\documentclass[useAMS,usenatbib]{mn2e}
\usepackage{epsf}
\usepackage{url}

\title[The morphology-density relation in MACSJ0717]{The morphology-density relation of galaxies around MACS\,J0717.5+3745\thanks{Based  on observations made with the NASA/ESA Hubble Space Telescope, obtained at the Space Telescope Science Institute, which is operated by the Association of Universities for Research in Astronomy, Inc., under NASA contract NASA 5-26555. These observations are associated with program GO-10420.}
}
\author[C.-J. Ma, \& H. Ebeling]{C.-J. Ma$^{1}$ and H. Ebeling$^{1}$\\
$^{1}$Institute for Astronomy, University of Hawaii, 2680 Woodlawn Drive, Honolulu, HI 96822, USA \\
}
\begin{document}

\date{}

\pagerange{\pageref{firstpage}--\pageref{lastpage}} \pubyear{2010}

\maketitle

\label{firstpage}

\begin{abstract}

We use an $18\arcmin \times 9\arcmin$ mosaic of HST/ACS images covering the entire large-scale structure around the X-ray luminous cluster MACSJ0717.5+3745 ($\rm z=0.545$) to study the morphology of galaxies at the cluster redshift. We find the global fraction of morphological types of galaxies to be consistent with results in the literature. In addition, we confirm the morphology-density relation of both early-type and late-type galaxies. Interestingly, we find that the fraction of lenticular galaxies (S0) also correlates with local galaxy density, in contrast to the findings of a study of the cores of 10 clusters at similar redshift by Dressler et al. (1997).
We suggest that this apparent inconsistency is due to differences in the spatial coverage around clusters, which is supported by the fact that the correlation disappears for S0s within a radius of 0.6R$_{200}$ of MACSJ0717.5+3745. We interpret this result as evidence of the morphology-density relation being caused by a combination of morphological transformation triggered by galaxy-galaxy interactions, and effects related to the formation and evolution of large-scale structure. In environments of low to intermediate density, where galaxy-galaxy interactions are frequent and efficient, the observed pronounced morphology-density relation for S0s reflects the density dependence of the interaction cross section. In clusters, however, the correlation disappears for S0s, as the much higher galaxy velocities in clusters not only lower the interaction cross section, but also cause a spatial redistribution of galaxies that all but destroys such a correlation. This argument does not hold for elliptical galaxies in clusters which, having formed much earlier, have settled into the large-scale cluster potential; hence the morphology-density relation for cluster ellipticals may reflect primarily the state of advanced dynamical relaxation of this population within the cluster rather than a causal link to the environment responsible for the morphological transformation of galaxies.

\end{abstract}


\section{Introduction}\label{sec:intro}

Our understanding of the morphological evolution of galaxies has advanced significantly since the discovery of the morphology-density relation \citep[T-$\Sigma$ relation hereafter,][]{dressler80,oemler74,melnick77} thirty years ago. It is generally accepted that the observations of more spirals and less S0s at higher redshift \citep[][but see also Andreon et al. 1998 and Holden et al. 2009 for a different view]{dressler97,fasano00,treu03,goto03,postman05,smith05,desai07,poggianti09b} imply that local S0s are the products of a morphological transformation. 

Various physical mechanisms may be responsible for this transformation, including galaxy-galaxy mergers \citep[][]{barnes91}, harassment \citep[][]{moore96}, ram-pressure stripping \citep[][]{gunn72}, and tidal distortion by the cluster potential \citep[][]{byrd90}. The efficiency of the first two mechanisms correlates with the local galaxy density but decreases dramatically with increasing relative velocities which lead to low galaxy-galaxy interaction cross sections. Hence, both mergers and harassment are likely to contribute significantly to the T-$\Sigma$ relation found in environments of low to intermediate density \citep[e.g.][]{goto03b}. In a high-density environment like clusters, however, the latter two mechanisms are more efficient. Regardless of its physical origin, a pronounced T-$\Sigma$ correlation is observed in the local universe for all morphological types \citep[][]{dressler80, dressler97, fasano00, thomas06}.

In recent times, the transformation of late-type to lenticular galaxies occurred in large numbers, as evidenced, for instance, by the actual dominance of S0s in clusters in the local universe \citep[][and references therein]{fasano00}. The temporal evolution (with redshift) of the morphological mix of galaxies in various environments remains unclear though. The gradual increase of the S0 population can, currently, be traced back only to redshifts z~$>0.5$; at higher redshift their fractional abundance appears to remain unchanged \citep[e.g.][]{desai07}. Possibly related, a T-$\Sigma$ correlation has not been observed for S0s in clusters at z~$\sim 1$ \citep{postman05} and will thus have to have been built up since z$\sim 1$. 

Studies of the T-$\Sigma$ relation of, specifically, S0s at intermediate redshift are few, compared to those of early-type and late-type galaxies. The classical work of \citet{dressler97} shows a flat T-$\Sigma$ relation of S0 galaxies in clusters at z~$\sim0.5$ in a study of galaxies in the central $0.4-0.8$~h$^{-1}$~Mpc of clusters. Recent observations \citep[e.g.][]{wilman09}, however, suggest that the formation of S0 galaxies may be most efficient in less dense environments, such as galaxy groups. Thus, no pronounced T-$\Sigma$ relation may be observed if the study is restricted to the galaxies within the core of clusters. 

We here present a wide-field study of galaxies around the massive cluster MACS\,J0717.5+3745 (z~$=0.545$) to examine the T-$\Sigma$ relation of S0s at intermediate redshift and to shed light on the impact of environment on the morphological transformation of galaxies.

Following an introduction of the target of our study in \S\ref{sec:m0717} and a brief description of our  data-reduction procedure in \S\ref{sec:data}, we describe our morphological classification scheme  in  \S\ref{sec:morphology}. In \S\ref{sec:morphology-density}, we present the T-$\Sigma$ relation. Finally, \S\ref{sec:discussion} discusses and summarizes our results.  Throughout this chapter, we adopt the concordance $\Lambda$CDM cosmology with $h_0=0.7$, $\Omega_{\lambda}=0.7$, $\Omega_m = 0.3$. Magnitudes are quoted in the AB system.

\begin{figure*}
\epsfxsize=0.9\textwidth
\epsffile{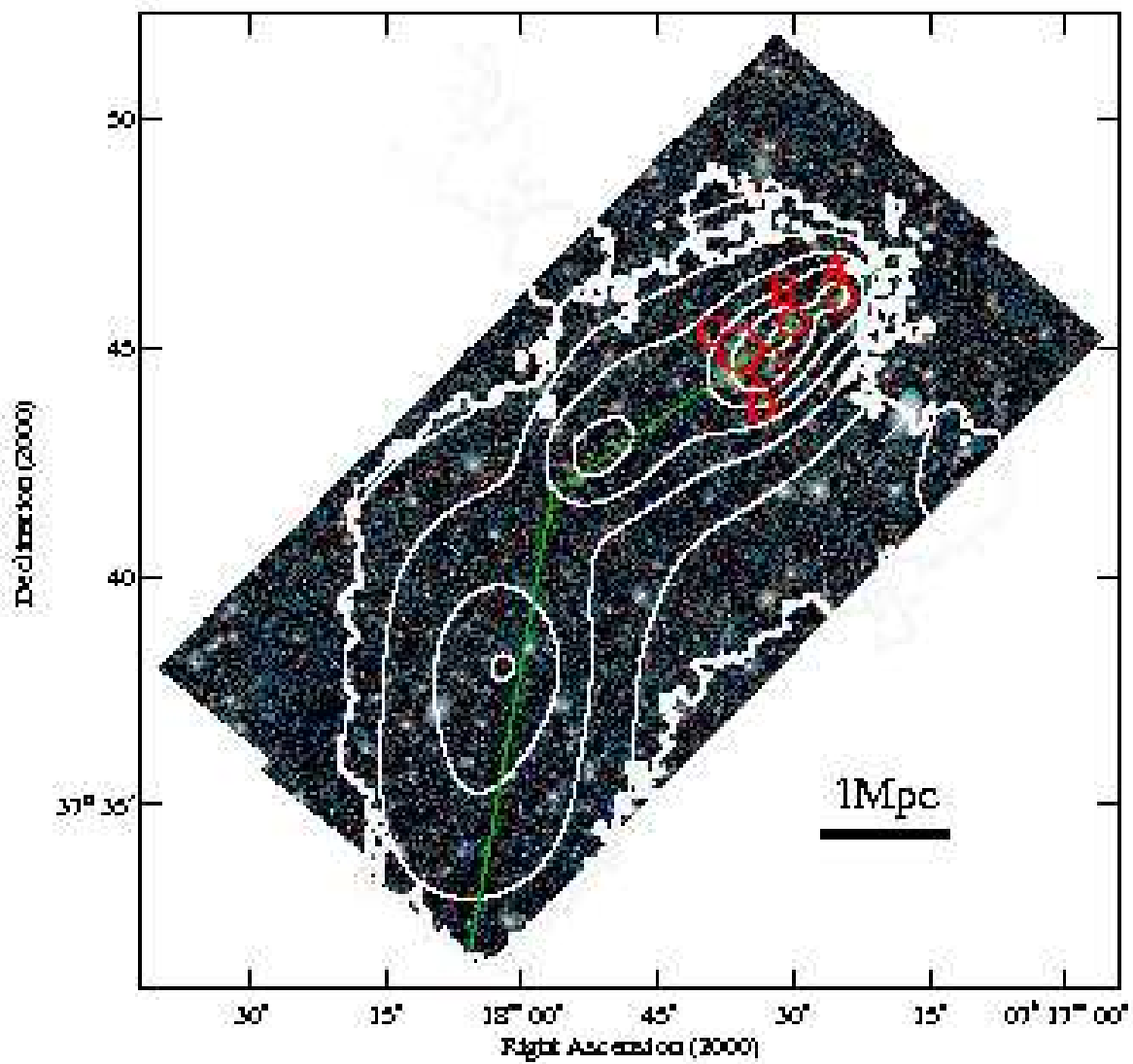}
\caption[Substructures near the core of MACSJ0717 and the filament]
{Area covered with HST/ACS images; overlaid are contours of the adaptively smoothed galaxy surface density (same as in \citet{ma08}). The red circles mark the locations of the merging components of the main cluster identified in \citet{ma09}; the green line schematically traces the filament.  \label{hstim}}
\end{figure*}

\begin{table*}
\centering  \caption{Number statistics of galaxies in MACSJ0717. The overall statistics for the different morphologies types are shown in the first line, together with the corresponding fractions (in parentheses).\label{number_galaxies}}
    \begin{tabular}{lcccc}
\hline 
Type & All     & S     & S0    & E  \\
\hline
all         & 657 & 333 ($0.51\pm 0.03$) & 126 ($0.19\pm 0.02$) & 198 ($0.30\pm 0.02$)\\ \hline
photometric $z$	   &  214     & 151  & 25   & 38    \\
spectroscopic $z$  &  443     & 182  & 101  & 160 \\
\cline{1-5}\vspace{-3mm} \\
absorption-line &  267     &  42  & 78   & 147  \\
E+A		&  16      &   9  &  6   &  1 \\
emission-line   & 160      & 131  & 17   & 12  \\
\hline
\end{tabular}

\end{table*}

\begin{figure*}
  \epsfxsize=1.0\textwidth 
  \epsffile{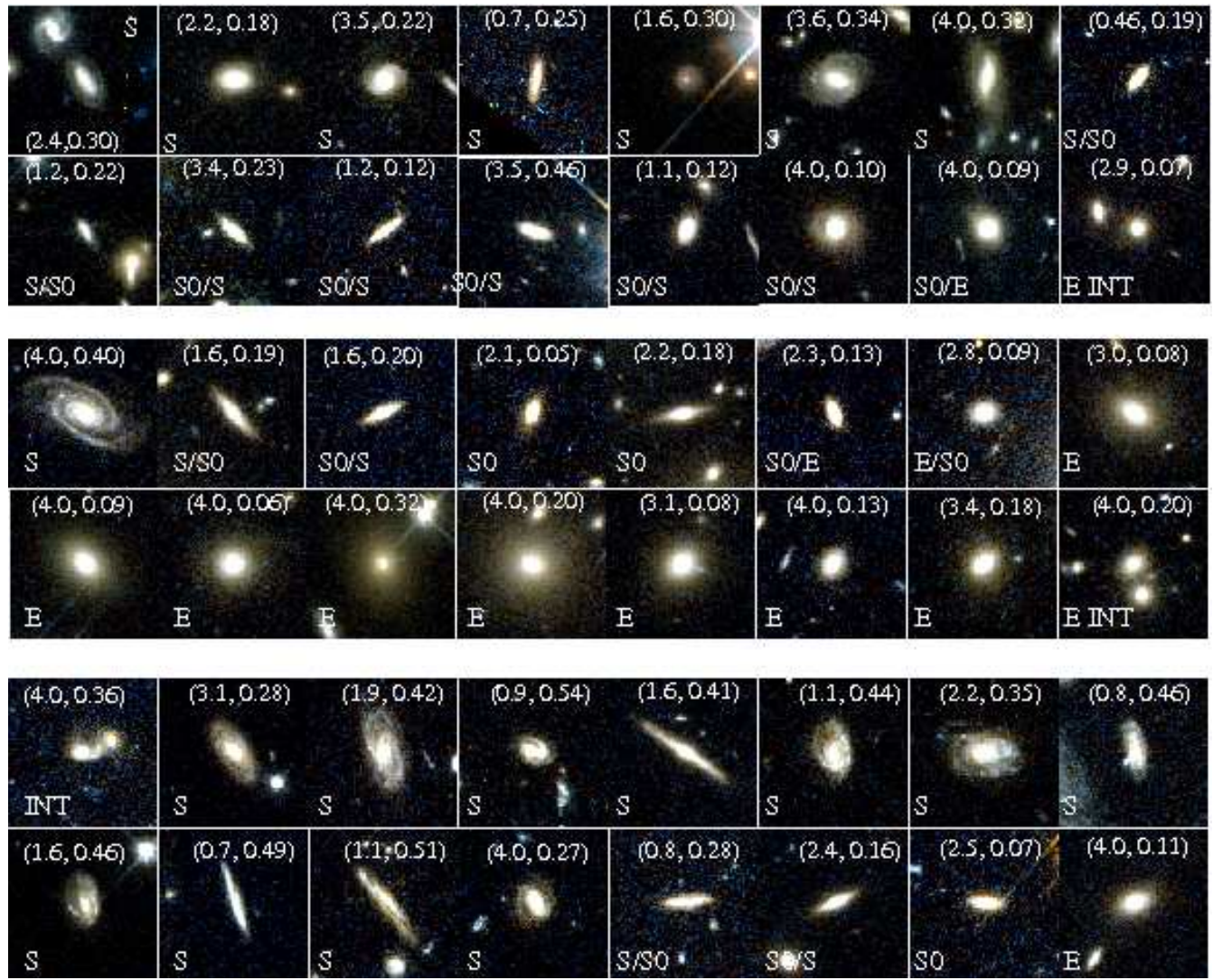}    
  \caption[ACS stamps of cluster galaxies in MACSJ0717]{Images (F606W and F814W) of selected cluster galaxies. The 16 cluster members within the ACS mosaic that have been classified as E+A galaxies by \citet{ma08} are shown in the top panel; the middle and bottom panels show examples of absorption-line and emission-line galaxies, respectively. The morphological type indicated in the lower left corner of each image is the result of a identification process that combines visual inspection and automatic classification as described in detail in \S\ref{sec:morphclass}.    The numbers shown in the parentheses in each panel are Sersic index (n) and Bumpiness (B), respectively. \label{example_stamps}}
\end{figure*}

\begin{figure*}
  \epsfxsize=0.49\textwidth 
  \epsffile{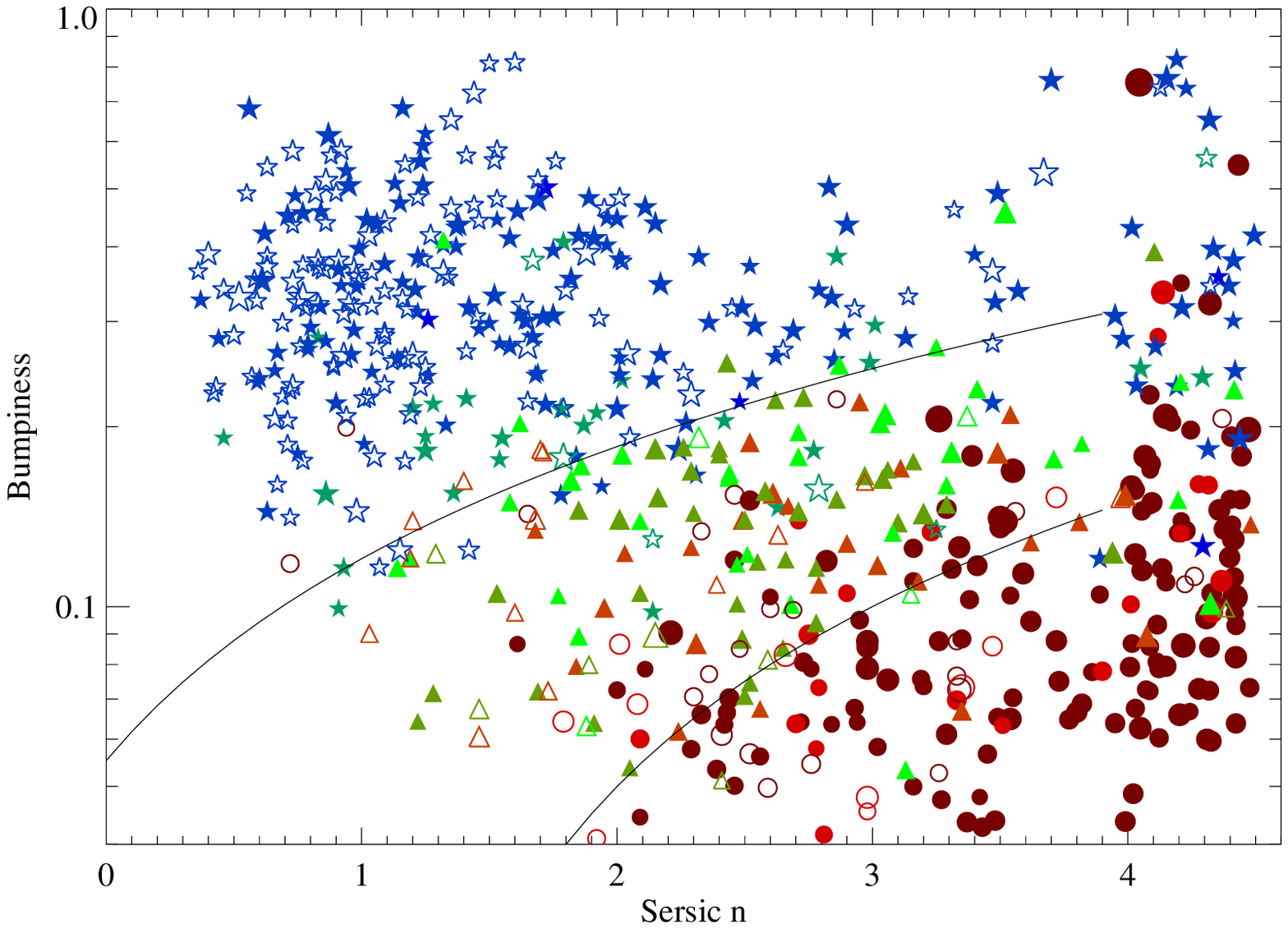}  
  \epsfxsize=0.49\textwidth 
  \epsffile{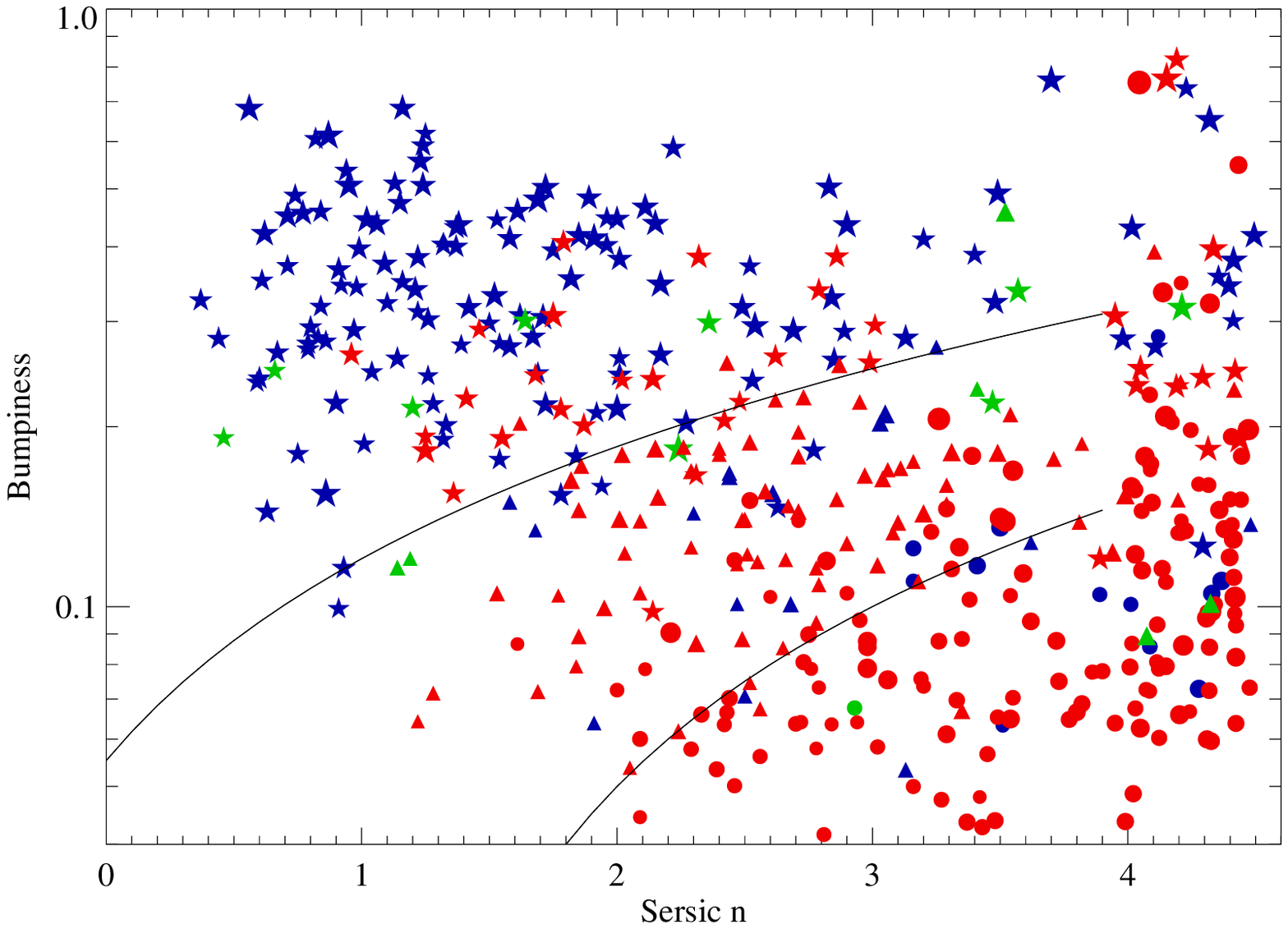}  
  \caption[B-n morphological classification of galaxies in MACSJ0717]{Distribution of Sersic indices and bumpiness parameters for MACSJ0717 cluster members. The symbols indicate morphological class: early-type (circles), lenticular (S0, triangles), and late-type galaxies (stars). The symbol size is proportional to the total R$_c$-band magnitude. In both panels, the two solid lines show the linear relations $B=0.065(n+0.8)$ and $B=0.05(n-1)$, which divide early- and late-type galaxies, and S0 and elliptical galaxies, respectively \citep{blakeslee06}. In the left panel, the shading indicates morphological sub-type (dark red represents E, red E/S0, orange S0/E, green S0, and so on). Filled symbols identify cluster members with spectroscopic redshifts; open symbols mark galaxies classified as cluster members based on their photometric redshift.  In the right panel, we show only cluster members that have spectroscopic data; here the color indicates the spectroscopic types: blue for emission-line, green for E+A, and red for absorption-line galaxies.\label{nB}}
\end{figure*}

\section{MACSJ0717.5+3745}\label{sec:m0717}

MACSJ0717.5+3745 (MACSJ0717 hereafter) is one of the most X-ray luminous clusters at z$>$0.5 in the MAssive Cluster Survey \citep[MACS][]{macs,ebeling07}\footnote{Global X-ray properties of  MACSJ0717 are summarized in \citet{ma08}}.  As discussed in \citet{ebeling04,ma08,ma09,bonafede09,vanweeren09}, it is a complex merging system connected to a linear large-scale structure and the only cluster to exhibit and obvious filament on Mpc scales in the sample of \citet{jeyhan08}. Within $1$~Mpc (projected radius) of the cluster core, studies of the density and temperature distribution of the intra-cluster medium (ICM) as well as of the galaxy distribution find the main cluster to be a triple merger \citep{ma09}. The system is a prime target for galaxy evolution since, in addition to the high-density cluster core, it comprises a wide range of environments on larger scales in the form of several X-ray detected satellite clusters/groups embedded in an almost linear (but slightly bent) filament extending almost 6 Mpc to the South-East of the main cluster (Fig.~\ref{hstim}).

\section{Observation and data reduction}\label{sec:data}

This is the fourth paper in a series on MACSJ0717, again using extensively the data presented in \citet{ebeling04}, \citet{ma08} and \citet{ma09}, i.e.\ a photometric catalogue in seven bands (BVR$_c$I$_c$z$^\prime$ obtained with the Suprime-Cam wide-field imager on Subaru \citep{suprimecam}, u$^*$ as observed with MEGACAM/CFHT, and JK$_s$ of WIRCAM/CFHT \citep{wircam} provenance), and galaxy spectra obtained primarily with the  DEep Imaging Multi-Object Spectrograph (DEIMOS) on Keck-II. For this work, we add to these data sets a mosaic of images in two passbands obtained with the Advanced Camera for Surveys (ACS) aboard the Hubble Space Telescope.

\subsection{Cluster galaxy catalogue}\label{sec:data_catalogue}

We use the catalogue of cluster galaxies as compiled by \citet{ma08}. It contains, first, all spectroscopically identified cluster members, defined as all galaxies with $\rm 0.522<z_{spec}<0.566$\footnote{The range of spectroscopic redshifts corresponds to $\rm z=z_{cl}\pm3\sigma_{\rm spec}$. The cluster redshift z$_{cl}=0.5446$ and velocity dispersion $\sigma_{\rm spec}=1612$ km s$^{-1}$ are determined from the redshift distribution of galaxies within the virial radius of the cluster \citep{ebeling07}.}, and second, all galaxies with photometric redshifts $\rm 0.48< z_{ph}<0.61$\footnote{The range of photometric redshifts is given by $\rm z_{ph}=z_{cl}\pm2\sigma_{ph}$, where $\sigma_{\rm ph}$ is the Gaussian scatter in the residuals between spectroscopic and photometric redshifts.}. The magnitude limit of the combined catalogue is set to $\rm m_{R_c}<22.5$. For reference, the absolute R$_{\rm c}$- and V-band magnitudes of galaxies at the magnitude limit of the catalogue are typically $\rm M_{R_c}\sim-20.5$ and $\rm M_{V}~\sim-20.2$. 

The distribution among several subclasses of the galaxies thus selected is summarized in Table.~\ref{number_galaxies}. In doing so, we split the overall sample not only according to morphological type, but also provide a breakdown into the three spectroscopic classes (absorption-, emission-line, and E+A galaxies)\footnote{The emission-line galaxies are defined as galaxies with detectable [OII] or $H_{\beta}$. For those galaxies with no [OII] and $H_{\beta}$ emission features, the absorption-line and E+A galaxies are distinguished using the equivalent width of $H_{\delta}$:  EW($H_{\delta}$)${<}4$\AA\, for the former, and  EW($H_{\delta}$)${>}4$\AA\, for the latter. }. Although an investigation into the correlation of morphological and spectroscopic types is not central to the work performed here, we present the respective statistics for use in, and comparison with, future studies.

\subsection{ACS images}\label{sec:data_acs}

MACSJ0717 was first observed with HST/ACS in April 2004 (PropID GO-9722, PI: Ebeling) and revisited in January 2005 (PropID GO-10420, PI: Ebeling) to cover the filamentary structure discovered in \citet{ebeling04}. The entire data set consists of 18 images in two filters (F814W and F606W) covering a contiguous area of $18\arcmin \times 9\arcmin$ (see Fig.~\ref{hstim}). At the position of each tile, the exposure time in the F814W passband was 4020 secs and 1980s in the F606W filter. The F814W image at the center of the cluster, obtained earlier for GO-9722, is slightly deeper with an exposure time of 4560 secs. 

The ACS data reduction is performed using the HAGGLeS pipeline (Marshall et al.\ 2010, in preparation; Schrabback et al. 2009), which employs MultiDrizzle\footnote{MultiDrizzle version 2.7.0} \citep{koekemoer02} for distortion correction, cosmic-ray rejection, and stacking. Individual frames at each mosaic position are drizzled together onto a $0.03\arcsec$ pixel grid \citep{schrabback09}. We then extract quadratic postage stamps of each cluster galaxy of size 8~half-light-radius on the side as estimated by SExtractor\citep[][]{bertin96}, but requiring a minimum size of $5\arcsec$ (see examples in Fig.~\ref{example_stamps}). For galaxies falling within more than one tile we select for our morphological analysis the image that is farther from the respective tile boundary. 

\section{Analysis of galaxy morphologies}\label{sec:morphology}

\subsection{Surface brightness fit and Bumpiness}\label{sec:data_galfit}

In order to quantify the morphology of galaxies within the entire MACSJ0717 complex, we fit the two-dimensional surface-brightness distribution of the selected galaxies in the F814W passband with a Sersic model and an additional sky-background component using GALFIT\footnote{http://users.obs.carnegiescience.edu/peng/work/galfit/galfit.html}\citep{galfit02}. This procedure is described in detail in \citet{ma10}. We here only note that the Sersic index $n$ is constrained to $0.2<n<4.0$, since allowing larger values of $n$ can cause the best-fit value of the effective radius to become unreasonably large without a significant improvement of the fit. Also, neighboring objects will only be fit simultaneously with additional Sersic models if they are connected to the target galaxy in the segmentation map generated by SExtractor. Otherwise, any neighboring objects appearing in the stamps will be simply masked out. 

Using the best-fit Sersic model, we can estimate the bumpiness parameter\footnote{
The bumpiness ($B$) is defined as the normalized residual of the surface-brightness distribution, $I$, and the fitted Sersic model, $S{(R_e,n)}$: 
$B = { \sqrt{ \langle\left[\, I - S{(R_e,n)} \right]_s^2\rangle - \langle\sigma_s^2\rangle} \over \langle S{(R_e,n)}\rangle}$, where $R_e$ is the effective radius, and $\sigma$ is the flux uncertainty of the observed surface-brightness image. We compute these averages over an annulus with an inner radius of two pixels from the galaxy centre and an outer radius of $2R_e$. The subscript $s$ indicates smoothing by a Gaussian function with FWHM$=0.085\arcsec$. }
 \citep[$B$; defined in][]{blakeslee06} for each galaxy, which quantifies the deviation of the smoothed Sersic model from the observed surface-brightness distribution. Since the $B$ values of galaxies with non-radial structures, such as spiral arms or tidal features, are larger than the values for galaxies which are well described by a smooth Sersic model, the bumpiness parameter is useful to distinguish between early- and late-type galaxies. We note, however, that, in practice, an early-type galaxy may be assigned a large $B$ value if its surface-brightness distribution cannot be fit well by a single Sersic model, but requires multiple Sersic-models of different effective radii. Thus, the $B$ value alone is not a robust morphological identifier.

\begin{figure}
  \epsfxsize=0.49\textwidth 
  \epsffile{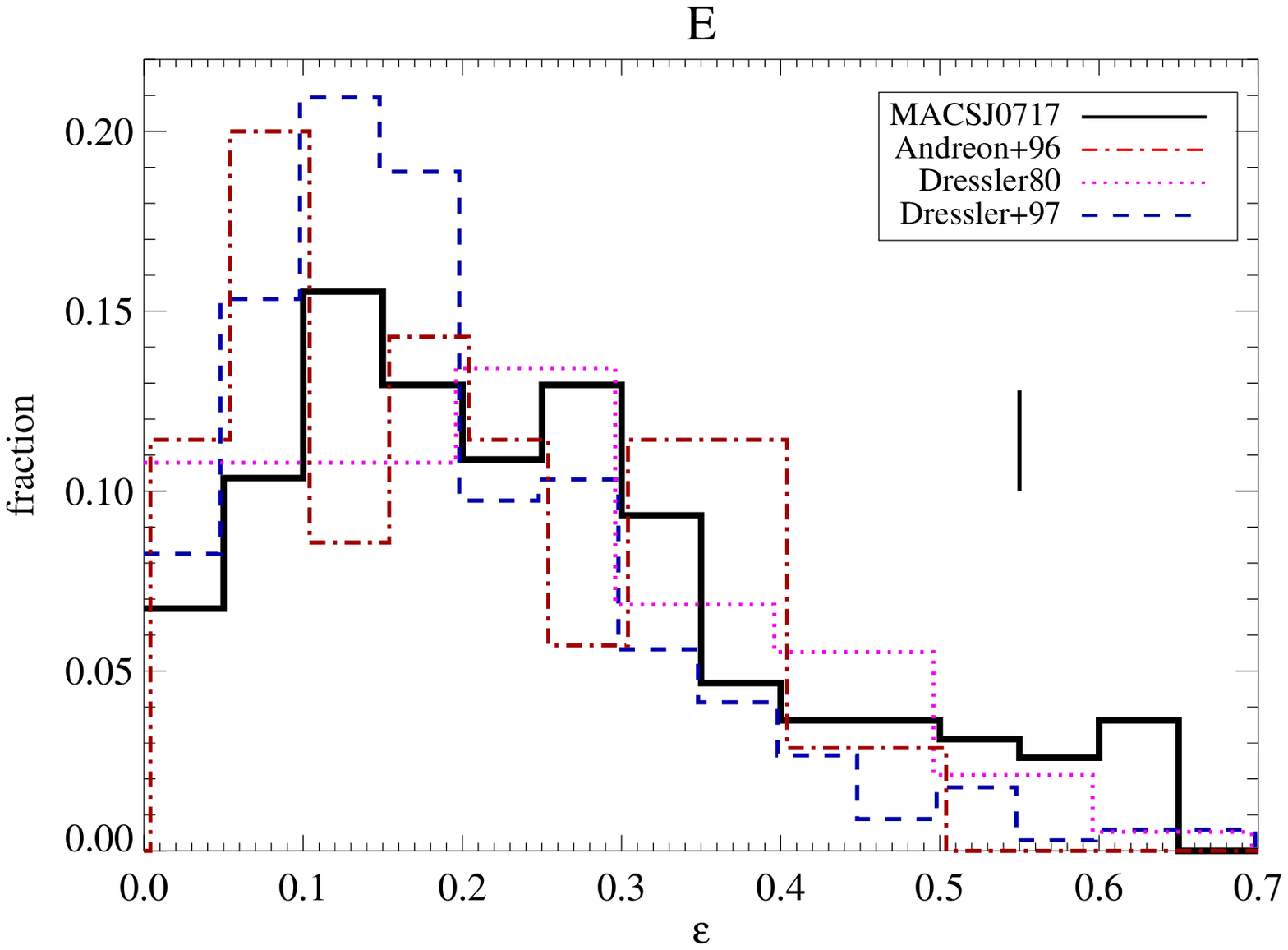}  
  \epsfxsize=0.49\textwidth 
  \epsffile{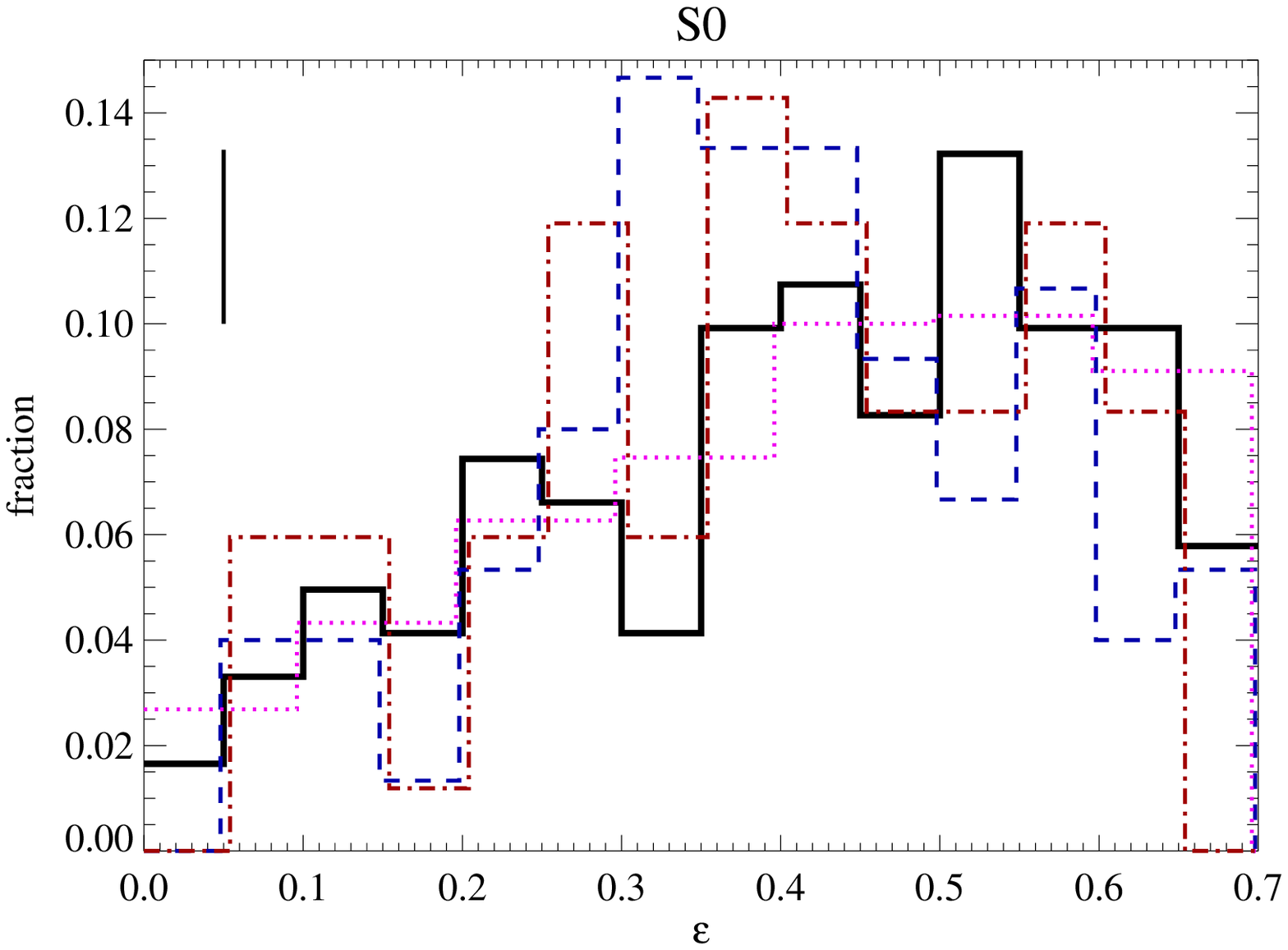}  

  \caption[Distribution of ellipticity of E and S0 of MACSJ0717]{Distribution of ellipticity of ellipticals (top panels) and S0s (bottom panels) in MACSJ0717 compared to results from the literature for both local clusters \citep{dressler80,andreon96}, and ten clusters at $0.37<z<0.56$ \citep{dressler97}. For reference, the vertical lines show the $1\sigma$ uncertainty for the highest bin of our own data. \label{ellipticity}}
\end{figure}

\subsection{Galaxy morphology}\label{sec:morphclass}

We follow the same classification scheme as used in \citet{ma10} to characterize the morphological type of cluster galaxies according to three major Hubble types, i.e. elliptical (E), spheroidal (S0), and late-types (S), where the late-type galaxies include all disk, interacting, and irregular galaxies. Galaxies that cannot be clearly assigned to one of these major types are categorized as sub-types, S0/E or E/S0, where the order indicates preference. When we calculate the fraction of morphological types, these sub-types, however, will be counted under their most probable major type. 

\citet{ma10} compare two classification methods, visual classification and the already mentioned Bumpiness-Sersic scheme \citep[$B$-$n$ method hereafter; see][]{blakeslee06, vanderwel07}, and find that both have their own advantages and disadvantages. The $B$-$n$ method provides a quantitative, operational definition of S0s, but sometimes fails for  early-type galaxies that cannot be described well by a single Sersic model. On the other hand, we can easily distinguish between early-type and late-type galaxies by eye, but the classification of S0 can be subjective. We, therefore, combine the two methods; the classification is primarily performed visually, and the $B$-$n$ method is used to characterize the visually ambiguous cases. In detail, the ACS postage stamps of individual galaxies are examined in random order and compared by eye with the templates shown in \citet{postman05}. In ambiguous cases,  galaxies will be classified as E/S0 if $B<0.05(n-1)$, S/S0 if $B>0.065(n+0.08)$, and S0/S or S0/E for values of $B$ in between the two division lines (Fig.~\ref{galxdensity}). 

Fig.~\ref{nB} shows the values of Bumpiness and Sersic index for all cluster galaxies within the ACS mosaic. Using the right panel of Fig.~\ref{nB}, and Table~\ref{number_galaxies}, we can compare the morphological and spectroscopic types identified in \citet{ma08}. We find that a large fraction of the E+A galaxies in MACSJ0717 are late-type galaxies, and that their Sersic indices are spread out over the entire range, in contrast to the E+A galaxies in MACSJ0025 \citep{ma10} which typically feature Sersic indices of $n\sim2$. In addition, we find 42 passive spirals \citep{moran07b,dressler99}, defined as absorption-line galaxies with late-type (S) morphology.

In Fig.~\ref{ellipticity}, we compare the ellipticity\footnote{Ellipticity is one free parameter in the Sersic-model fit described in \S\ref{sec:data_galfit}.}  distribution of S0s and ellipticals in MACSJ0717 to that obtained for other clusters in previous studies. For both S0s and ellipticals these distributions are, in general, consistent\footnote{A two-sided Kolmogorov-Smirnov test finds the differences to be significant at about the 1-sigma level.} with the ones from the literatures \citep{dressler80, andreon96, dressler97, lane07} which strengthens our confidence in the validity of our morphological classification.  


\section{Morphology-density relation} \label{sec:morphology-density}

\begin{figure}
\epsfxsize=0.45\textwidth 
\epsffile{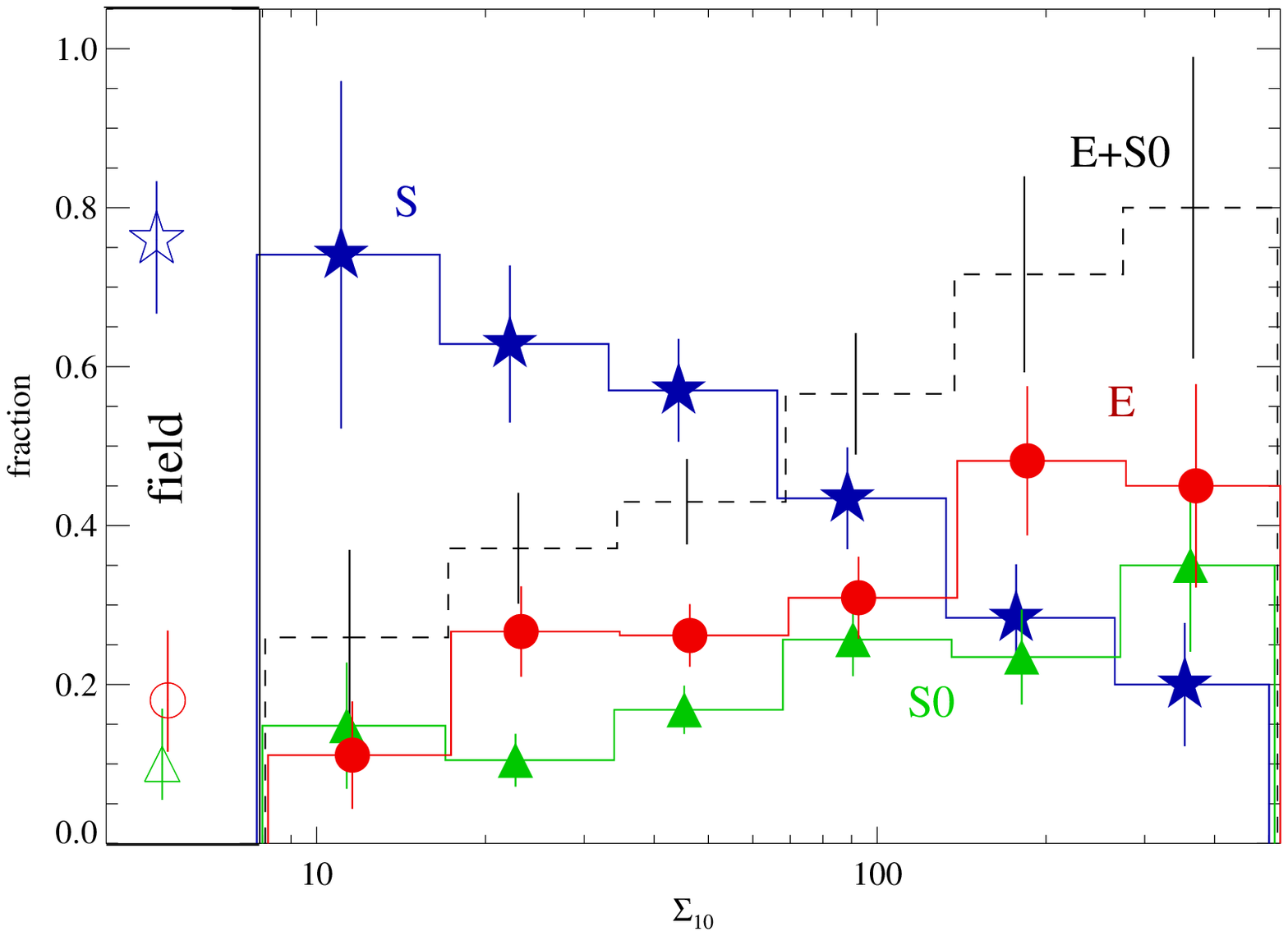}
\epsfxsize=0.45\textwidth 
\epsffile{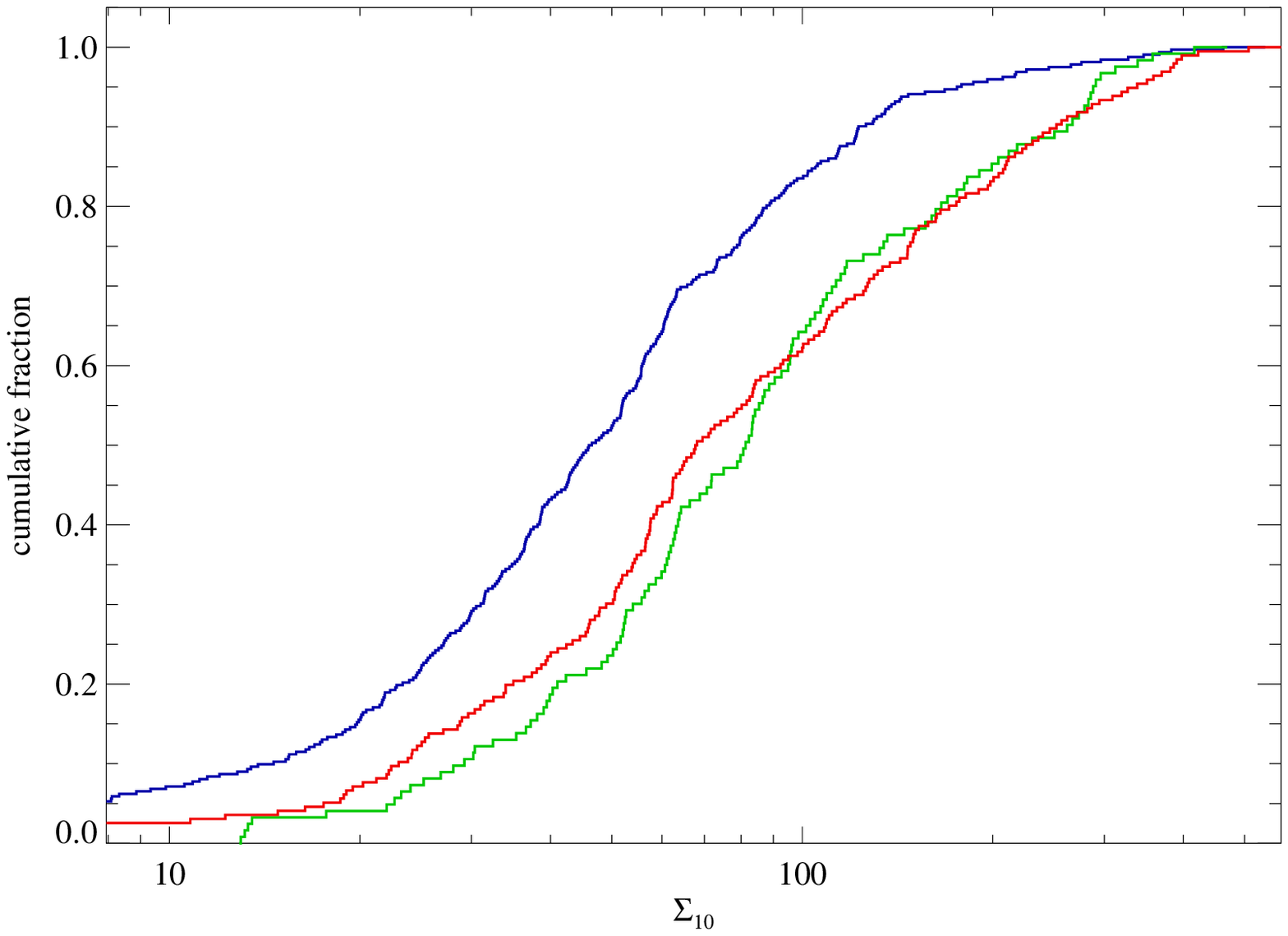}
\caption[T-$\Sigma$ relation of galaxies at MACSJ0717 redshift]{T-$\Sigma$ relation of galaxies around MACSJ0717. In the top panel, the fraction of different morphological classes are plotted with respect to $\Sigma_{10}$, our estimator of the projected galaxy density. The three points on the far left mark the fraction of three morphological classes for field galaxies (defined in \S\ref{sec:morphology-density}). In the bottom panel, the normalized cumulative distribution of different morphological classes are shown.\label{galxdensity}}
\end{figure}

To derive the T-$\Sigma$ relation, we adopt the standard local galaxy-density estimator $\Sigma_{10}$, defined as the surface density out to the 10th nearest neighbor galaxy. $\Sigma_{10}$ is calculated using all galaxies with photometric redshifts consistent with the cluster redshift (\S\ref{sec:data_catalogue}), and with magnitudes R$_{c}<24$ in \citet{ma08}.

We find the fraction of late-type galaxies (S) to decrease monotonically with local galaxy density, whereas the fraction of early-type galaxies (E+S0) shows the opposite trend (Fig.~\ref{galxdensity}). The mix of morphological types at the highest galaxy densities is qualitatively consistent with the results for other clusters at similar redshift \citep[e.g.][]{dressler97,poggianti09b}. We note that a detailed comparison should account for selection effects introduces by, e.g., differences in the magnitude limits or the definition of morphological types and cluster membership. In addition, the morphological mix of cluster galaxies at the lowest galaxy densities is consistent with that of field galaxies, which are selected from our spectroscopic catalogues by imposing $\rm 0.44<z_{spec}<0.48$ and $\rm 0.61<z_{spec}<0.65$.

Interestingly, the fraction of S0 galaxies at the cluster redshift increases with local galaxy density, a trend observed in local clusters \citep[][]{dressler80,thomas06}, but not observed before in clusters at $z\ga 0.5$ \citep[][]{dressler97, postman05}. 
A comparison of the normalized cumulative distributions for the three morphological classes, shown in the bottom panel of Fig.~\ref{galxdensity}, further confirms the similarity between the trends of S0 and E galaxies. A two-sided Kolmogorov-Smirnov test finds that the difference between these two distributions to be significant about the 1-sigma level, and thus consistent with both distributions being drawn from an identical parent population. To our knowledge, this is the first detection of the T-$\Sigma$ relation of S0s at z~$\sim0.5$. 

\subsection{Morphology-radius relation}\label{sec:morphology-radial}

\begin{figure}
\epsfxsize=0.45\textwidth
 \epsffile{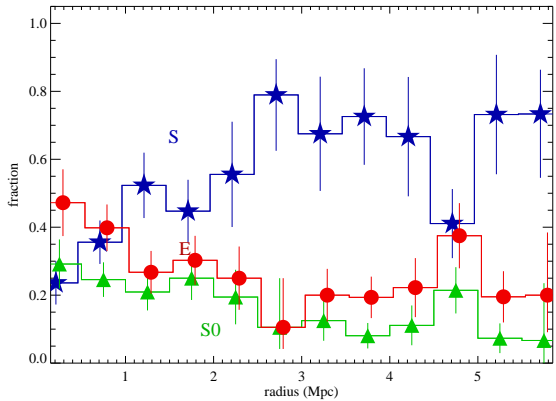}
\caption[Distribution of galaxy morphology with respect to cluster radial distance]{Distribution of galaxies of different
  morphology classes with cluster-centric distance. The symbols are the same as in Fig.~\ref{galxdensity}. Some symbols are shifted slightly by $\pm0.02$~Mpc for
  clarity. \label{galxradial}}
\end{figure}

The correlation between galaxy morphology and cluster-centric distance is not as fundamental as the T-$\Sigma$ relation \citep[e.g.][]{treu03,thomas06,lane07} because of its dependence on spherical symmetry. Confirming this caveat, we find that the morphology-radius relation\footnote{We define the cluster center as the location of the peak of the X-ray surface brightness \citep{ma09}.} of galaxies at  the redshift of MACSJ0717 (shown in Fig.~\ref{galxradial}) is not as clean as the T-$\Sigma$ relation. The fractions of E and S0 galaxies only increase within $\sim2$~Mpc radius around MACSJ0717 but then stay flat until a radius of about $4-5$~Mpc, which marks the location of the secondary galaxy-density peak near the endpoint of the filament (Fig.~\ref{hstim}).

\subsection{Correlation with global cluster properties and cluster redshift}\label{sec:morphology-global}

The existence of correlations between galaxy morphology and global cluster properties is more controversial than the T-$\Sigma$ relation. Several studies find correlations between the morphology of galaxies and the velocity dispersion and X-ray luminosity of their host clusters \citep[e.g.][]{poggianti09b,desai07}. For MACSJ0717, we find global morphological fractions of $0.51\pm 0.03$ (S), $0.19\pm 0.02$ (S0), $0.30\pm 0.02$ (E) (see also Table~\ref{number_galaxies}). Within 0.6$R_{\rm 200}$ (the radius used in the study by \citet{poggianti09b}) the balance shifts, unsurprisingly, toward a higher fraction of early-type galaxies: $0.28\pm 0.05$ (S), $0.29\pm 0.05$ (S0), $0.42\pm 0.06$ (E). These values are similar to the ones observed in MACSJ0025.4--1222 ($z{=}0.584$) in work by \citet{ma10} who find morphological fractions of $0.35\pm 0.04$ (S), $0.29\pm 0.04$ (S0), and $0.36\pm 0.04$ (E). Both sets of morphological fractions are consistent with the results compiled by \citet{poggianti09b} for a large sample of clusters covering a wide range of velocity dispersions and X-ray luminosities. Our measurements suggest, however, a flattening of the correlations of early- and late-type fractions with cluster X-ray luminosity for extremely luminous systems (see also Fig.~7 of \citep{ma10}). 

As mentioned in \S\ref{sec:intro}, the fraction of morphological types evolves strongly with cluster redshift. For lenticulars, it increases from a value of roughly 20\% at redshift 0.4 to 1, to about 50\% in the local Universe \citep{dressler97,fasano00,postman05,desai07}. Our value of $0.29\pm 0.05$ for the S0 fraction within 0.6$R_{\rm 200}$ of MACSJ0717 ($z{=}0.545$) is thus slightly high in comparison, but still consistent with all results in the literature. Thanks to the high optical richness of our target cluster, ours constitutes one of the statistically most robust measurements of morphological fractions at any redshift.

\section{Summary and interpretation}\label{sec:discussion}

Our primary results are twofold: first, we find the global fractions of morphological types among the galaxy population of MACSJ0717 ($z{=}0.545$) to be consistent with those of other clusters at similar redshift \citep[][and references therein]{wilman09}. Second, our wide-field study of the galaxy population in and around MACSJ0717 shows that a T-$\Sigma$ correlation for lenticular galaxies exists already at $z{=}0.545$. 

\begin{figure}
\epsfxsize=0.45\textwidth 
\epsffile{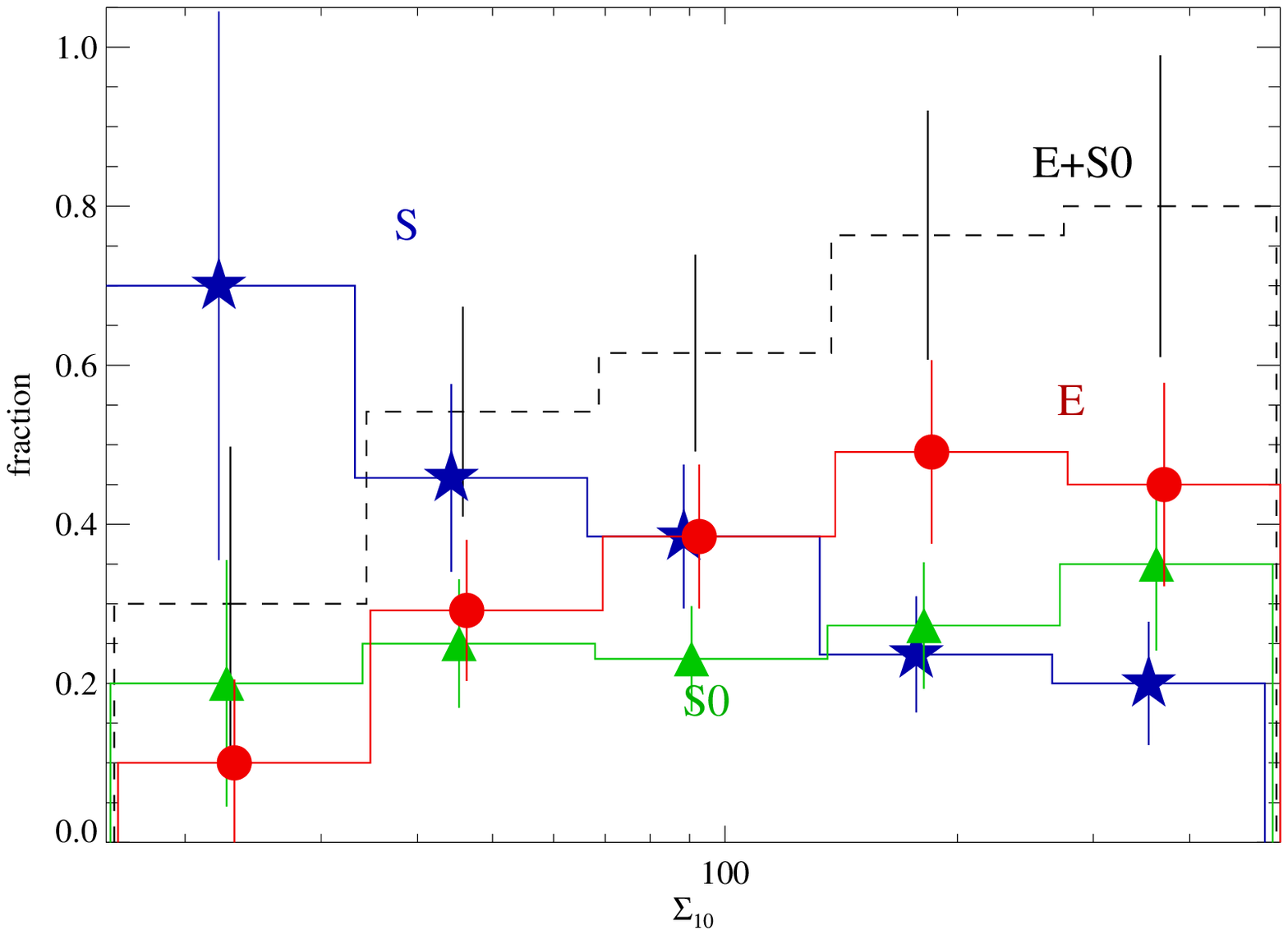}
\epsfxsize=0.45\textwidth 
\epsffile{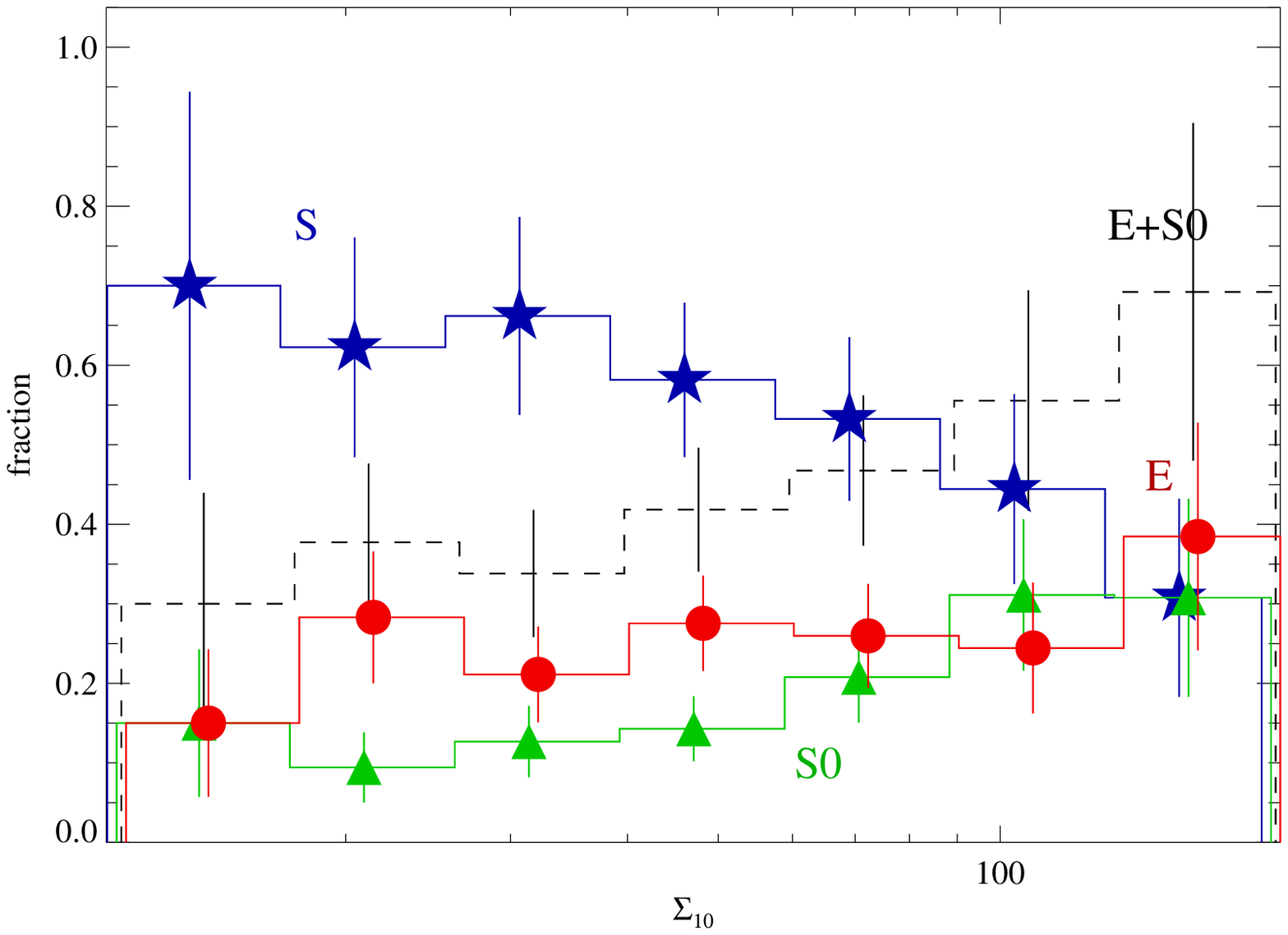}
\caption[Morphology-density relation of galaxies within $\rm 0.6R_{200}$]{Same as Fig.~\ref{galxdensity}, but for galaxies with $\rm r<0.6R_{200}$ (Top) and $\rm r>0.6R_{200}$ (Bottom). For reference, a linear regression shows the slopes of S0 fractions as a function of log($\Sigma_{10}$) are $0.10\pm0.13$ (Top) and $0.20\pm0.09$ (Bottom). \label{galxdensity_in}}
\end{figure}

At face value, our second result appears to be different with the findings of \citet{dressler97} who observe no significant correlation between S0 fraction and galaxy density in a study of the cores of 10 clusters at comparable redshift. A simple linear regression finds ($0.16\pm0.05$)) for the slope of the T-S relation of S0s in MACSJ0717 (i.e. a significant correlation), compared to ($0.049\pm0.048$) for the intermediate-redshift sample of Dressler et al. (1997). The discrepancy can be explained, however, by the different areal coverage of the respective surveys. To allow a fair comparison, we show in Fig.~\ref{galxdensity_in} the T-$\Sigma$ relations for MACSJ0717 using only galaxies within (top) and beyond (bottom) a radius of $0.6R_{200}\sim 1.2$~Mpc from the cluster core. We find the correlation apparent in Fig.~\ref{galxdensity} to be generated mainly in the cluster outskirts, with little contribution from the core region, consistent with earlier work. The different behaviour of elliptical and lenticular galaxies in the two panels of  Fig.~\ref{galxdensity_in} also argues strongly against our findings being due to, or significantly affected by, errors or biases in our morphological classification.

The strong dependence of the T-$\Sigma$ correlation on cluster-centric radius gives us a valuable clue to the physical mechanisms underlying the morphological transformation. One explanation for the observed flat distribution within cluster cores could be that the transformation mechanisms at work are the cluster-specific ones (ram-pressure stripping and tidal disruption caused by the cluster potential, \S\ref{sec:intro}) which do not depend directly on local galaxy density. This explanation, however, cannot account for the pronounced T-$\Sigma$ relation detected in our study once the survey area is enlarged to include the cluster outskirts. The efficiency of galaxy mergers and harassment, on the other hand, increases with galaxy density and decreases with relative velocity; hence, galaxy-galaxy interactions are highly implausible causes of morphological transformation in the very cores of clusters. By contrast, the efficiency of moderate-velocity galaxy-galaxy encounters for morphological transformation is evidenced by many studies of galaxy evolution in intermediate-density environment \citep{helsdon03,wilman09}. Specifically, \citet{moran07b} demonstrates that suitable environments include the outskirts of clusters. As shown in bottom panel of Fig.~\ref{galxdensity_in} this is indeed the environment that contributes the bulk of the observed T-$\Sigma$ correlation.

A correlation between morphological fractions and local galaxy density could, however, still be expected in environments between the aforementioned extremes, i.e., closer to the cluster center but far enough from the very core for relative galaxy velocities to be low enough to be conducive to merging and harassment \citep[][]{moore99,mihos04}. No convincing correlation of this kind is, however, present in the top panel of Fig.~\ref{galxdensity_in}. We propose an explanation for this observation in the following paragraph.

For the purposes of the following argument we separate the lenticular galaxies found in clusters at moderate to high redshift into two populations, but note that this distinction need not imply different physical origins\footnote{We will explore and attempt to quantify he relative efficiencies of cluster-specific mechanisms and of galaxy-galaxy interactions in a future study.}. At and beyond the outskirts of clusters, galaxies transformed as the result of galaxy-galaxy interactions at low relative velocity dominate the S0 population, and hence their numbers correlate with local galaxy density. Lenticular galaxies found in the cores of clusters, however, were largely accreted from the low-density environment at larger cluster-centric distances where they formed. 
This cluster-core population does not exhibit  a correlation of the S0 fraction with local galaxy density because we observe and recognize these galaxies as lenticulars only after the morphological transformation has been largely completed. This process takes of the order of a few Gyr \citep[][]{kodama01b}, which is larger than the cluster crossing time. Hence the location of these members of the second population of S0s is not representative of the environment that initially triggered their morphological transformation and no T-$\Sigma$ correlation is observed\footnote{This argument remains valid if a fraction of the lenticulars in cluster cores were in fact created {\it in situ} through cluster-specific mechanisms, since the distribution of this sub-population still would not correlate with local galaxy density.}.

In the local universe this picture is slightly altered, as the cluster lenticulars created in the processes discussed above will have settled into virial equilibrium, meaning their spatial distribution will now trace the gravitational potential of the cluster. Intriguingly, a direct consequence of this argument is that the T-$\Sigma$ relation observed for S0s in local clusters \citep[e.g.][]{dressler80} does not establish a causal link between these two quantities, but is mainly the result if a fortuitous mix of a population that traces the cluster potential (ellipticals and old lenticulars) with a second population that does not (newly created S0s and spirals still being accreted from the field). For elliptical galaxies which may have been created as early as $z\sim 2$ \citep{vandokkum01} the same argument holds also for clusters out to $z\sim 1$, i.e., even at moderate to high redshift the observed pronounced T-$\Sigma$ correlation tells us little about the cause of the morphological transformation from late- to early-type galaxies. 

Finally, we emphasize that we do not suggest that the T-$\Sigma$ relation of S0s in the local universe in general is unrelated to the density dependence of the morphological transformation of galaxies -- it is only in clusters that much of the observed correlation is likely to be spurious. In the low- to intermediate-density regime, this relation  still constitutes one of the most convincing pieces of evidence of density-dependent galaxy evolution. However, when interpreting this relation in cluster cores we need to bear in mind that the galaxy density at the observed location of an S0 is not indicative of the density of the environment within which the morphological transformation originated.

Our discovery of a significant T-$\Sigma$ correlation in this study demonstrates that the large-scale environment of MACSJ0717 is conducive to the creation of lenticular galaxies via galaxy-galaxy interactions already at $z{=}0.55$, which leaves us with a conundrum. The fact that no significant evolution in the fraction of any morphological galaxy type is observed at $z>0.35$ would only be understandable, if the physical mechanisms discussed in \S\ref{sec:intro} were not efficient at intermediate to high redshift. Our finding, however, shows that at least mergers, and possibly also harassment, are at work, a result that does not come unexpected in view of the fact that the galaxy environment at $z=0.4-1$ is not dramatically different from the one encountered at, e.g., $z\sim 0.1$. Why then, does the global fraction of lenticular galaxies remain constant all the way from $z\sim 0.35$ to $z\sim 1$? A partial answer may be provided by a hypothesis advanced by \citet{smith05} who argue that a significant fraction of S0s forms so early that it cannot be explained by the transformation of infalling late-type galaxies. While this argument may explain why the global fraction of S0s does not decrease to essentially zero at high redshift, it fails to account for the absence of any apparent evolution at $z>0.35$. 

Future large-scale studies of massive galaxy clusters at $z>0.3$ may provide the best statistics to accurately establish both the redshift at which the T-$\Sigma$ correlation finally disappears and the global morphological fraction of S0s which continue to play a pivotal role in our understanding of galaxy evolution.

\section*{Acknowledgments}
 
 We thank Phil Marshall and Tim Schrabback for providing state-of-the-art HST/ACS data reduction via the HAGGLeS legacy project, and Ian Smail for very helpful comments on a draft version of this paper. CJM and HE gratefully acknowledge financial support from STScI grants GO-09720 and GO-10420. 


\end{document}